\documentclass[letterpaper]{jpsj-suppl}
\usepackage{ctable}

\title{Updated Analysis of the Mass of the H Dibaryon from Lattice QCD}

\author{P.~E.~\textsc{Shanahan}, A.~W.~\textsc{Thomas} and R.~D.~\textsc{Young}}

\inst{ARC Centre of Excellence in Particle Physics at the Terascale and CSSM, \\School of Chemistry and Physics, University of Adelaide, Adelaide SA 5005, Australia}

\email{phiala.shanahan@adelaide.edu.au}

\recdate{July 5, 2013}

\abst{

Recent lattice QCD calculations from the HAL and NPLQCD Collaborations have reported evidence for the existence of a bound state with strangeness $-$2 and baryon number 2 at quark masses somewhat higher than the physical values. 
A controlled chiral extrapolation of these lattice results to the physical point suggested that the state, identified with the famed H dibaryon, is most likely slightly unbound (by $13 \pm 14$~MeV) with respect to the $\Lambda-\Lambda$ threshold. 
We report the results of an updated analysis which finds the H unbound by $26 \pm 11$~MeV.

Apart from the insight it would give us into how QCD is realized in Nature, the H is of great interest because of its potential implications for the equation of state of dense matter and studies of neutron stars. It may also explain the enhancement above the $\Lambda-\Lambda$ threshold already reported experimentally. It is clearly of great importance that the latter be pursued in experiments at the new J-PARC facility.

}

\kword{nuclear strangeness, H dibaryon}

\begin{document}
\maketitle

\section{Introduction}

Our understanding of quantum chromodynamics has been challenged for decades by the apparent absence of multiquark states. The outstanding candidate for such a state has been the H dibaryon, ever since it was proposed by Jaffe in 1977~\cite{Jaffe:1976yi} that this strangeness $-2$ state should be very deeply bound with respect to the $\Lambda-\Lambda$ threshold. Since then, a variety of models have been used to calculate the mass of the H, with results varying widely; while some predict a deeply-bound or light H, others predict a lightly bound or even unbound state~\cite{Sakai,Carames,Mulders:1982da}. Extensive experimental efforts to find the particle~\cite{Yoon:2007aq,Takahashi:2001nm,Stotzer:1997vr,Iijima:1992pp} have led to the conclusion that it does not appear to be bound.

We consider the extremely exciting possibility that the H dibaryon, which had been abandoned by many, may indeed be almost bound with respect to the $\Lambda - \Lambda$ threshold. 
This possibility was raised recently by the NPLQCD and HAL collaborations~\cite{Inoue:2010es,Beane:2010hg}, which found the H bound by tens of MeV at the unphysically large light quark masses where the calculations were performed.

A chiral extrapolation to the physical point, assuming that the H is indeed a genuine multi-quark state and not a quasi-molecular state like the deuteron, found that at the physical quark masses the H is most likely slightly unbound, by $13 \pm 14$~MeV~\cite{Shanahan:2011su,Thomas:2011cg}.

We report here the results of an updated analysis, taking into account revised and additional results for the binding of the H from the HAL and NPLQCD Collaborations~\cite{Inoue:2011ai,Beane:2011iw}.
We also describe further the limitations of this calculation, and discuss the experimental context and cosmological implications of the results. Our updated analysis finds the H dibaryon at $26 \pm 11$~MeV above the $\Lambda-\Lambda$ threshold.

\section{Updated calculation of H dibaryon mass}

Given evidence from lattice QCD for a bound H dibaryon at light quark masses somewhat larger than the physical values, it is imperative to perform an extrapolation of these results to the physical point. 
Shanahan {\it et al.}~\cite{Shanahan:2011su,Thomas:2011cg} presented a chiral extrapolation of the HAL and QCDSF Collaboration lattice results under the assumption that the H is a compact 6-quark state (not a molecular state like the deuteron). In that case, as the H is an SU(3)-singlet, its mass may be expressed as
\begin{equation}
M_H = M_H^{(0)}-\sigma_H \left( \frac{m_\pi^2}{2}+m_K^2 \right) + \delta M_H^{(3/2)},
\end{equation}
where $\delta M_H^{(3/2)}$ represents quantum corrections associated with the chiral loops involving the pseudo-Goldstone bosons $\pi$, $K$, $\eta$. The loops, which include those with both octet and decuplet baryon intermediate states, are regularized using the `finite range regularization' technique~\cite{Leinweber:2003dg,Young:2002ib,Young:2002cj,Donoghue:1998bs,Leinweber:1998ej}. This method incorporates the correct leading non-analytic chiral behaviour in the extrapolation, while being applicable over a larger light quark mass range than traditional dimensional regularization.

A chiral extrapolation for the binding of the H dibaryon, defined as
\begin{equation}
B_H=2M_\Lambda-M_H,
\label{eq:binding}
\end{equation}
may be written, given the extrapolation for $M_H$ and the usual form of chiral extrapolation for the $\Lambda$ baryon mass $M_\Lambda$.

As there are few lattice results available for the binding of the H, the chiral extrapolation is constrained by simultaneously fitting the lattice data for $B_H$ and a set of PACS-CS Collaboration lattice results for the octet baryon masses~\cite{Aoki:2008sm}. 

\begin{figure}[f]
\begin{center}
\includegraphics[width=0.77\columnwidth]{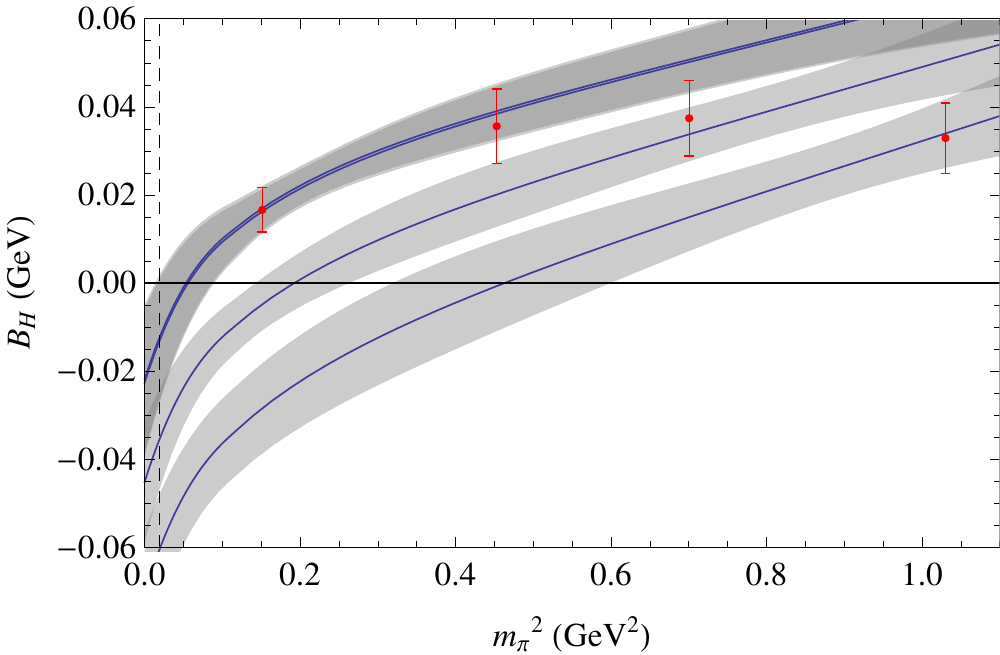}
\caption{Binding of the H dibaryon against pion mass squared, resulting from the chiral fit described in Ref.~\cite{Shanahan:2011su}, for several values of the strange quark mass at which the HAL QCD and NPLQCD lattice simulations were carried out (2011).}
\label{fig:orig}
\end{center}
\end{figure}

\begin{figure}[f]
\begin{center}
\includegraphics[width=0.77\columnwidth]{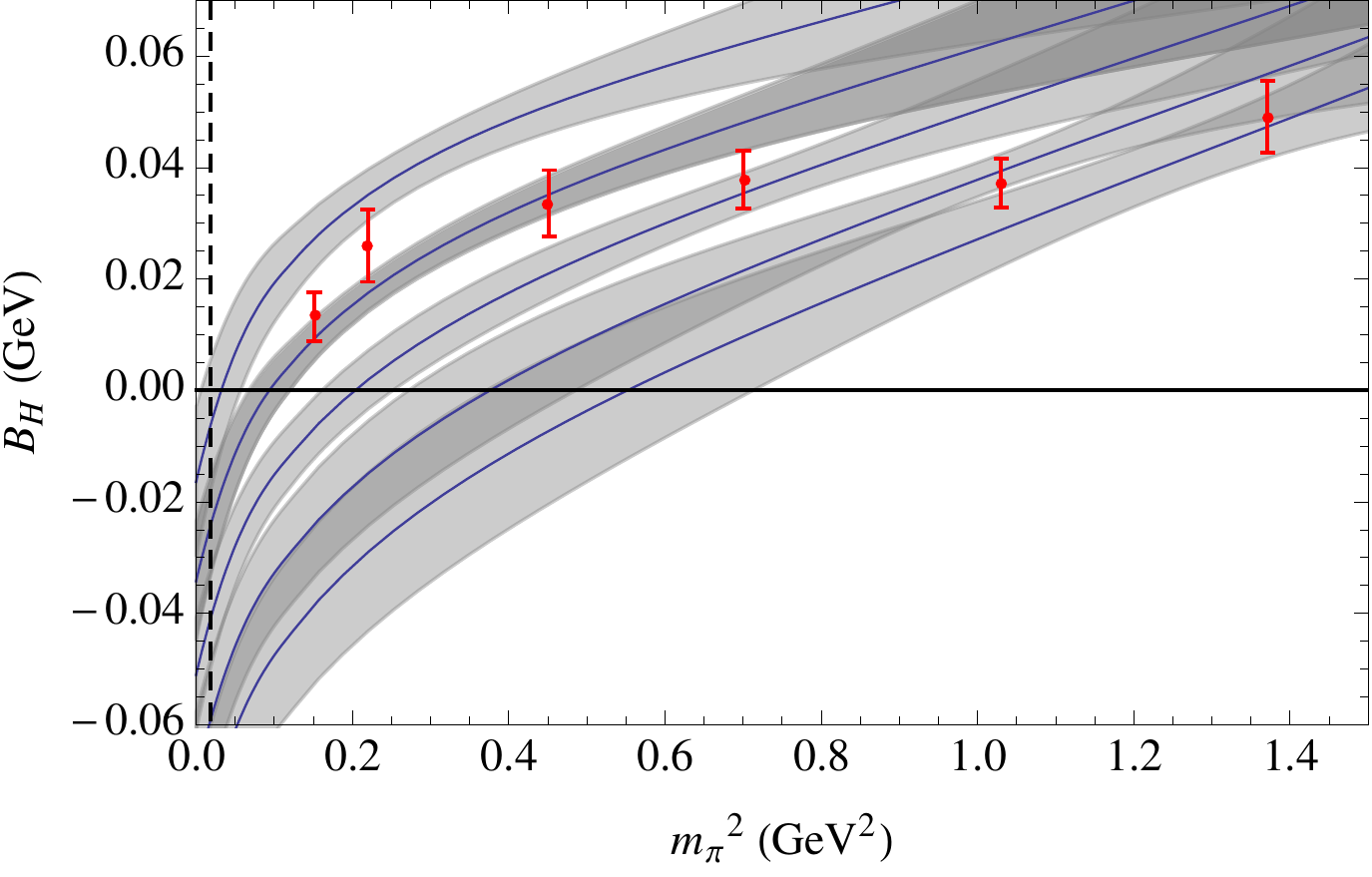}
\caption{Updated plot of binding of the H. All points and errors have been updated as described in the text, and two new HAL Collaboration (SU(3)-symmetric) points have been included.}
\label{fig:updated}
\end{center}
\end{figure}

The lattice results for $B_H$ have recently been updated by each collaboration:
\begin{itemize}
\item The HAL QCD Collaboration used hadron-hadron potentials extracted from the time-dependent Nambu-Bethe-Salpeter wave function to extract $B_H$. All points are calculated at the SU(3)-symmetric point, $m_l=m_s$, on a single volume\cite{Inoue:2010es}. Updated values and two new mass points are given in Ref.~\cite{Inoue:2011ai}.
\item The NPLQCD Collaboration used L\"uscher's method using finite volume scattering phase shifts. They calculated one mass point on four volumes and performed the infinite-volume extrapolation~\cite{Beane:2010hg}. Updated values are given in Ref.~\cite{Beane:2011iw}.
\end{itemize}
Repeating the analysis described in detail in Refs.~\cite{Shanahan:2011su,Thomas:2011cg} with all lattice data for $B_H$ updated yields the $H$ unbound by 26$\pm$11~MeV, consistent with the previous results. The quality of fit for each (original and updated) data set is shown in Figs. \ref{fig:orig} and \ref{fig:updated}, and best-fit values for all fit parameters are given in Table~\ref{tab:orig}.

The most significant limitation of the calculation is that the effect of a change in quark mass on the coupling of the H to any of the open baryon-baryon channels could not be included in the analysis, as there is no information on this from lattice QCD. It can be argued that (as it is an SU(3)-singlet) the H couples most strongly to the $\Sigma-\Sigma$ and $\Xi-N$ channels, which one expects to contribute an attraction varying slowly with quark mass. It will nevertheless be of significant interest to see the results of coupled-channel $2+1$-flavor calculations being undertaken by the HAL Collaboration, as well as to include the preliminary light $m_\pi=230$~MeV lattice data point from the NPLQCD Collaboration~\cite{Beane:2011zpa}.

It is important to note that the calculation described above makes the assumption that the H dibaryon is a single hadron with 6 quarks, as proposed by Jaffe in the context of his 1977 bag-model calculation. A second common definition of `dibaryon' is as a baryon-baryon molecule.
%BUT might expect significant variation in moving away from SU(3) symmetric point? Most of the lattice results (all of the HAL points) take $m_l=m_s$ so could be significant...?
Haidenbauer and Mei\ss ner~\cite{Haidenbauer:2011za} perform a chiral perturbation theory analysis with the assumption that the H is a loosely bound $BB$ state, like the deuteron, with a similar binding energy to that of the deuteron. They take into account the shifts of the $\Lambda-\Lambda$, $\Sigma-\Sigma$, $\Xi-N$ thresholds with quark mass. They find results quantitatively similar to our analysis, finding a resonance at 21~MeV based on the NPLQCD lattice results, and no surviving H from the HAL lattice results.

\begin{center}
\begin{table}[tb]
\centering
\begin{tabular}{c  r@{.}l r@{.}l r@{.}l r@{.}l r@{.}l}
\toprule
 & \multicolumn{2}{c}{$\Lambda$ (GeV)} &  \multicolumn{2}{c}{$M_0$~(GeV)} &  \multicolumn{2}{c}{$\alpha$ (GeV$^{-1}$)} &  \multicolumn{2}{c}{$\beta$ (GeV$^{-1}$)} &  \multicolumn{2}{c}{$\sigma$ (GeV$^{-1}$)} \\ \midrule
original analysis & 1&02(6) & 0&86(4) & -1&71(12) & -1&20(10) & -0&51(5) \\ 
updated analysis & 0&98(8) & 0&90(2) & -1&54(6) & -1&01(6) & -0&44(3) \\ \midrule[\heavyrulewidth]
&   \multicolumn{2}{c}{$B_0$~(GeV)} &  \multicolumn{2}{c}{$\sigma_B$ (GeV$^{-1}$)}&  \multicolumn{2}{c}{$C_H$~(GeV$^{-2}$)} &  \multicolumn{2}{c}{$\chi^2/\textrm{d.o.f.}$}\\ \midrule
 & 0&019(4) & -2&36(20) & 5&65(9) & 0&48 \\
 & 0&041(25) & -1&66(25) & 5&23(29) & 0&64 \\ \bottomrule
\end{tabular}
\caption{Values of the fit parameters for the octet and H dibaryon 
data corresponding to the 
fits shown in Figs.~\ref{fig:orig} and \ref{fig:updated}.}
\label{tab:orig}
\end{table}
\end{center}

\section{Cosmological implications of an H dibaryon}

While the original interest in the H dibaryon was for its own sake as a candidate exotic particle, it was quickly noted that its existence would have significant consequences for models and descriptions of neutron stars~\cite{Schaffner-Bielich,Witten:1984rs}.

The existence (or absence) of hyperons is thought to affect:
\begin{itemize}
\item The mass-radius relation and maximum mass of compact stars;
\item The cooling of neutron stars;
\item The stability with regard to the emission of gravitational waves from rotating neutron stars.
\end{itemize}
In particular, the equation of state for compact stars is softened at high densities by the appearance of a dibaryon condensate, and can result in a mass plateau for these stars~\cite{Carames}. 
%See also [arXiv:astro-ph/9803232] for a discussion of the constraints neutrons stars can give for the mass of the H.

While most neutron star models which incorporate an H dibaryon assume a bound H, it must not be forgotten that the effect of dense nuclear matter could significantly change the structure of such a particle. In particular, an unbound free H near the $\Lambda-\Lambda$ threshold could acquire binding energy in a nucleus~\cite{Yamada:2000sh}.
Given a threshold H dibaryon, it is possible that the energetically favourable compression of H-matter could lead to the formation of uniform strange matter and thereby provide a pathway for a neutron star to convert into a strange quark star.

\section{Experiment}

In the light of the results reported above it is of significant interest to experimentally explore the near-threshold region of phase space for the H dibaryon. In this section we briefly describe the current experimental programs searching for the H.

\subsubsection{J-PARC}

The observation of the sequential decay of double hyperfragments provides the only known constraint for the binding of the H. Depending on the H mass, the ground state of an $S=-2$ hypernucleus will contain either an H or two $\Lambda$ hyperons. If $M_H < (2M_\Lambda - B_{\Lambda\Lambda})$, where $B_{\Lambda\Lambda}$ is the binding energy of the hypernucleus, the doubly strange nucleus can decay into an H and a daughter nucleus. Thus $B_{\Lambda\Lambda}$ can be used to provide a bound on $B_H$.
Few decay events have been observed to date. The NAGARA event (sequential decay of $_{\Lambda\Lambda}^6$He)~\cite{Takahashi:2001nm} gives the most stringent constraint on the binding of the H dibaryon, $B_H <6.91$~MeV.

J-PARC will perform a systematic study of doubly-strange hypernuclei, with ten times the statistics of the similar experiment KEK-PS E522. More than 3300 $\Lambda\Lambda$-nuclei are expected to be detected by means of $\Xi$-capture reactions using different target nuclei (C, N,O)~\cite{Yoon:2007aq}. This uses the double strangeness and charm exchange reaction:
\begin{equation}
K^-+p \rightarrow K^+ + \Xi^-.
\end{equation}
The $\Xi^-$ is brought to rest in the emulsion; nuclear emulsions are the only detection procedure with enough resolution to measure the short tracks produced by double hypernuclei. The mass resolution will be 1~MeV, compared to the 5~MeV mass resolution of the earlier KEK experiment. It is expected that this experimental effort will have sufficient statistics to resolve the long-held problem of the existence of the H dibaryon~\cite{Shyam:2012fp}.

The KEK-PS E224 and E522 experiments~\cite{Yoon:2007aq,Ahn:1998fj} saw enhancements in the $\Lambda-\Lambda$ spectrum perhaps providing a hint of an unbound H. This peak will be searched for in the J-PARC results. The difficult question is whether the observed bump is the signal of a resonance (i.e., the H) or can be explained simply by some non-resonant enhancement arising from an attractive final state interaction.

%and search for the bound H-dibaryon by its weak decays to answer the long-standing question about the existence of the H-dibaryon.

\subsection{PANDA, FAIR}

The PANDA experiment at FAIR will use antiproton-induced reactions for the formation of single and double $\Lambda$-hypernuclei~\cite{Wiedner:2011mf}.
This is complementary to the experiment at J PARC as the production of hypermatter, induced by heavy ions and antibaryons, in nuclear matter covers a broader region in baryon density, single-particle energy, and isospin asymmetry space, while hypernuclear structure provides information around saturation density only~\cite{Saito}.

\section{Conclusion}

We have updated a chiral perturbation theory extrapolation of lattice results for the binding of the H dibaryon. While at large light quark masses there is evidence for a bound H from lattice QCD, we find that chiral physics leads to a more rapid decrease of the mass of the $\Lambda$ than for the H as $m_\pi$ approaches its physical value. As a result one must conclude that at the physical values of the quark masses the H dibaryon is most likely unbound. Our estimate, including the effect of correlations between all the fit parameters, is that the H is unbound by $26 \pm 11$~MeV  at the physical point. 
This result is of particular significance taken in the context of the current high-statistics experimental efforts at J-PARC and FAIR to discover a possible H particle or resonance.

\section{Acknowledgements}

This work was supported by the University of Adelaide and the Australian Research Council through through the ARC Centre of Excellence for Particle Physics at the Terascale and grants FL0992247 (AWT) and DP110101265 (RDY) and FT120100821 (RDY).


\begin{thebibliography}{9}
%\cite{Jaffe:1976yi}
\bibitem{Jaffe:1976yi}
R.~L.~Jaffe,
%``Perhaps a Stable Dihyperon,''
Phys.\ Rev.\ Lett.\  {\bf 38}, 195-198 (1977).
%
%\cite{Sakai,Carames}
\bibitem{Sakai}
T.~Sakai, K.~Shimizu, K.~Yazaki,
Prog.\ Theor.\ Phys.\ Suppl.\ {\bf137} 121-145 (2000). 

\bibitem{Carames}
%THE RENEWED CHALLENGE OF THE H DIBARYON
T.~F.~Caram\'es, A.~Valcarce,
Int.\ J.\ Mod.\ Phys.\ {\bf E22}, 1330004 (2013). 


%\cite{Mulders:1982da}
\bibitem{Mulders:1982da}
P.~J.~Mulders, A.~W.~Thomas,
%``Pionic Corrections And Multi - Quark Bags,''
J.\ Phys.\ G {\bf G9}, 1159 (1983).


%\cite{Yoon:2007aq,Takahashi:2001nm,Stotzer:1997vr,Iijima:1992pp}
\bibitem{Yoon:2007aq}
C.~J.~Yoon, H.~Akikawa, K.~Aoki, Y.~Fukao, H.~Funahashi, M.~Hayata,
K.~Imai, K.~Miwa {\it et al.},
%``Search for the H-dibaryon resonance in C-12 (K-, K+ Lambda Lambda
%X),''
Phys.\ Rev.\  {\bf C75}, 022201 (2007).
%    
%\cite{Takahashi:2001nm}
\bibitem{Takahashi:2001nm}
H.~Takahashi, J.~K.~Ahn, H.~Akikawa, S.~Aoki, K.~Arai, S.~Y.~Bahk,
K.~M.~Baik, B.~Bassalleck {\it et al.},
%``Observation of a (Lambda Lambda)He-6 double hypernucleus,''
Phys.\ Rev.\ Lett.\  {\bf 87}, 212502 (2001).
%
%\cite{Stotzer:1997vr}
\bibitem{Stotzer:1997vr}
R.~W.~Stotzer {\it et al.} [ BNL E836 Collaboration ],
%``Search for H dibaryon in He-3 (K-, k+) Hn,''
Phys.\ Rev.\ Lett.\  {\bf 78}, 3646-3649 (1997).
%
%\cite{Iijima:1992pp}
\bibitem{Iijima:1992pp}
T.~Iijima, H.~Funahashi, S.~Hirata, M.~Ieiri, I.~Imai, T.~Ishigami,
Y.~Itow, K.~Kikuchi {\it et al.},
%``(K-, K+) reaction on nuclear targets at P(K-) = 1.65-GeV/c,''
Nucl.\ Phys.\  {\bf A546}, 588-606 (1992).
%



%\cite{Inoue:2010es,Beane:2010hg}
\bibitem{Inoue:2010es}
T.~Inoue {\it et al.} [HAL QCD Collaboration],
%``Bound H-dibaryon in Flavor SU(3) Limit of Lattice QCD,''
Phys.\ Rev.\ Lett.\  {\bf 106}, 162002 (2011).
%[arXiv:1012.5928 [hep-lat]].
%
%\cite{Beane:2010hg}
\bibitem{Beane:2010hg}
S.~R.~Beane {\it et al.} [NPLQCD Collaboration],
%``Evidence for a Bound H-dibaryon from Lattice QCD,''
Phys.\ Rev.\ Lett.\  {\bf 106}, 162001 (2011).
%[arXiv:1012.3812 [hep-lat]].
%

%\cite{Shanahan:2011su,Thomas:2011cg}
\bibitem{Shanahan:2011su} 
 P.~E.~Shanahan, A.~W.~Thomas and R.~D.~Young,
  %``Mass of the H-dibaryon,''
 Phys.\ Rev.\ Lett.\  {\bf 107}, 092004 (2011).
 %[arXiv:1106.2851 [nucl-th]].

%\cite{Thomas:2011cg}
\bibitem{Thomas:2011cg} 
 A.~W.~Thomas, P.~E.~Shanahan and R.~D.~Young,
  %``Strange quarks and lattice QCD,''
 Few Body Syst.\  {\bf 54}, 123 (2013).



%\cite{Inoue:2011ai,Beane:2011iw}
\bibitem{Inoue:2011ai} 
  T.~Inoue {\it et al.}  [HAL QCD Collaboration],
  %``Two-Baryon Potentials and H-Dibaryon from 3-flavor Lattice QCD Simulations,''
  Nucl.\ Phys.\ A {\bf 881}, 28 (2012).

%\cite{Beane:2011iw}
\bibitem{Beane:2011iw} 
  S.~R.~Beane {\it et al.}  [NPLQCD Collaboration],
  %``The Deuteron and Exotic Two-Body Bound States from Lattice QCD,''
  Phys.\ Rev.\ D {\bf 85}, 054511 (2012).
%  [arXiv:1109.2889 [hep-lat]].



%\cite{Leinweber:2003dg,Young:2002ib,Young:2002cj}
\bibitem{Leinweber:2003dg}
D.~B.~Leinweber, A.~W.~Thomas, R.~D.~Young,
%``Physical nucleon properties from lattice QCD,''
Phys.\ Rev.\ Lett.\  {\bf 92}, 242002 (2004).
%[hep-lat/0302020].
%
%\cite{Young:2002ib}
\bibitem{Young:2002ib}
R.~D.~Young, D.~B.~Leinweber, A.~W.~Thomas,
%``Convergence of chiral effective field theory,''
Prog.\ Part.\ Nucl.\ Phys.\  {\bf 50}, 399-417 (2003).
%[hep-lat/0212031].
%
%\cite{Young:2002cj}
\bibitem{Young:2002cj}
R.~D.~Young, D.~B.~Leinweber, A.~W.~Thomas, S.~V.~Wright,
%``Chiral analysis of quenched baryon masses,''
Phys.\ Rev.\  {\bf D66}, 094507 (2002).
%[hep-lat/0205017].

%\cite{Donoghue:1998bs}
\bibitem{Donoghue:1998bs}
J.~F.~Donoghue, B.~R.~Holstein, B.~Borasoy,
%``SU(3) baryon chiral perturbation theory and long distance regularization,''
Phys.\ Rev.\  {\bf D59}, 036002 (1999).
%[arXiv:hep-ph/9804281 [hep-ph]].

%\cite{Leinweber:1998ej}
\bibitem{Leinweber:1998ej}
D.~B.~Leinweber, D.~-H.~Lu, A.~W.~Thomas,
%``Nucleon magnetic moments beyond the perturbative chiral regime,''
Phys.\ Rev.\  {\bf D60}, 034014 (1999).
%[hep-lat/9810005].
%


%\cite{Aoki:2008sm}
\bibitem{Aoki:2008sm}
S.~Aoki {\it et al.} [ PACS-CS Collaboration ],
%``2+1 Flavor Lattice QCD toward the Physical Point,''
Phys.\ Rev.\  {\bf D79}, 034503 (2009).
%[arXiv:0807.1661 [hep-lat]].

%\cite{Beane:2011zpa}
\bibitem{Beane:2011zpa} 
  S.~R.~Beane {\it et al.},
  %``Present Constraints on the H-dibaryon at the Physical Point from Lattice QCD,''
  Mod.\ Phys.\ Lett.\ A {\bf 26}, 2587 (2011).


%\cite{Haidenbauer:2011za}
\bibitem{Haidenbauer:2011za} 
  J.~Haidenbauer and U.~G.~Meissner,
  %``Exotic bound states of two baryons in light of chiral effective field theory,''
  Nucl.\ Phys.\ A {\bf 881}, 44 (2012).




%
%\cite{Beane:2011xf}
\bibitem{Beane:2011xf}
S.~R.~Beane, E.~Chang, W.~Detmold, B.~Joo, H.~W.~Lin, T.~C.~Luu,
K.~Orginos, A.~Parreno {\it et al.},
%``Present Constraints on the H-dibaryon at the Physical Point from
%Lattice QCD,''
[arXiv:1103.2821 [hep-lat]].
%

%

\bibitem{Schaffner-Bielich} 
J.~Schaffner-Bielich, 
%"Strangeness in Compact Stars ",
Nucl.\ Phys.\ {\bf A835}, 279 (2010).

%\cite{Witten:1984rs}
\bibitem{Witten:1984rs} 
  E.~Witten,
  %``Cosmic Separation of Phases,''
  Phys.\ Rev.\ D {\bf 30}, 272 (1984).
  %%CITATION = PHRVA,D30,272;%%

%\cite{Yamada:2000sh}
\bibitem{Yamada:2000sh} 
  T.~Yamada and C.~Nakamoto,
  %``Structure of light S=-2 nuclei and hyperon mixing,''
  Phys.\ Rev.\ C {\bf 62}, 034319 (2000).
  %%CITATION = PHRVA,C62,034319;%%
%

%\cite{Shyam:2012fp}
\bibitem{Shyam:2012fp} 
  R.~Shyam, O.~Scholten and A.~W.~Thomas,
  %``Production of the H dibaryon via the (K^-,K^+) reaction on a $^{12}$C target,''
  arXiv:1211.0775 [hep-ph].
  %%CITATION = ARXIV:1211.0775;%%

%\cite{Ahn:1998fj}
\bibitem{Ahn:1998fj} 
  J.~K.~Ahn {\it et al.}  [KEK-PS E224 Collaboration],
  %``Enhanced Lambda Lambda production near threshold in the C-12(K-,K+) reaction,''
  Phys.\ Lett.\ B {\bf 444}, 267 (1998).
%
%\cite{Wiedner:2011mf}
\bibitem{Wiedner:2011mf} 
  U.~Wiedner,
  %``Future Prospects for Hadron Physics at PANDA,''
  Prog.\ Part.\ Nucl.\ Phys.\  {\bf 66}, 477 (2011)
  [arXiv:1104.3961 [hep-ex]].
  %%CITATION = ARXIV:1104.3961;%%

%\cite{Wiedner:2011mf}
\bibitem{Saito} 
  T.~R.~Saito {\it et al.},
  %``Future Prospects for Hadron Physics at PANDA,''
  Nucl.\ Phys.\  {\bf A835}, 110 (2010).

\end{thebibliography}
\end{document}